\documentclass[reprint,pra,aps]{revtex4-1}
\usepackage[colorlinks=true,urlcolor=blue,anchorcolor=blue,%
citecolor=blue,filecolor=blue,linkcolor=blue,menucolor=blue,%
linktocpage=true,pdfproducer=medialab]{hyperref}
\usepackage{braket}
\usepackage{bm}
\usepackage{amssymb,amsmath,amsthm}
\usepackage{mathrsfs}
\usepackage{xcolor}
\usepackage{appendix}
\usepackage{graphicx}
\usepackage{tikz}
\usepackage{dcolumn}
\usepackage{bm}
\usepackage{multirow}
\usepackage{float}
\usepackage[normalem]{ulem}
\definecolor{purple}{rgb}{1,0,1}
\definecolor{lime}{HTML}{A6CE39} 

\newcommand{\orcidicon}{%
	\begin{tikzpicture}
	\draw[lime, fill=lime] (0,0) 
		circle [radius=0.16] 
		node[white] {{\fontfamily{qag}\selectfont \tiny ID}};
	\draw[white, fill=white] (-0.0625,0.095) 
		circle [radius=0.007];
	\end{tikzpicture}
	\hspace{-3mm}
}
\newcommand\orcidDel{{\href{https://orcid.org/0000-0003-4158-202X}{\orcidicon}}}
\newcommand\orcidMatt{{\href{https://orcid.org/0000-0003-1088-6485}{\orcidicon}}}
\begin{document}
\title{Quantum Blockchain Using Entanglement in Time}

\author{Del Rajan\orcidDel{} {\sf and} Matt Visser\orcidMatt{}}

\affiliation{
School of Mathematics and Statistics, Victoria University of Wellington, \\
PO Box 600, Wellington 6140, New Zealand
}
\email{del.rajan@sms.vuw.ac.nz}
\email{matt.visser@sms.vuw.ac.nz}

\begin{abstract}\noindent
We propose a  conceptual design for a quantum blockchain.  Our method involves encoding the blockchain into a temporal GHZ (Greenberger--Horne--Zeilinger) state of photons that do not simultaneously coexist. It is shown that the entanglement in time, as opposed to an entanglement in space, provides the crucial quantum advantage.  All the subcomponents of this system have already been shown to be experimentally realized. Furthermore, our encoding procedure can be interpreted as nonclassically  influencing the past.	 

\smallskip
\noindent
{\sc Pacs:} 03.67.Bg, 03.67.Dd, 03.67.Hk,03.67.Mn


\smallskip
\noindent
{\sc arXiv:}  1804.05979 [quant-ph]

\smallskip
\noindent
{\sc Dated:} 17 April 2018; 18 April 2019; \LaTeX-ed \today

\smallskip
\noindent
{\sc Published:} Quantum Reports {\bf 1 \# 1 } (2019) 3--11.
\url{https://doi.org/10.3390/quantum1010002}

\end{abstract}

\maketitle
\section{Introduction}
\label{S:intro}
\def\N{{\mathbb{N}}}
\def\Z{{\mathbb{Z}}}
\def\P{{\mathbb{P}}}
\def\implies{\Longrightarrow}
\newtheorem{conjecture}{Conjecture}{}

Entanglement is an intrinsically quantum effect that involves nonclassical correlations, usually between spatially separated quantum systems~\cite{nielsen2010quantum}.  This phenomenon was described by Einstein as ``spooky action at a distance'', and yet it forms the basis of nearly all quantum information platforms, such as quantum computers and quantum networks.  
In particular, quantum networks distribute quantum information between any two nodes on the network~\cite{castelvecchi2017quantum,castelvecchi2018quantum}.  This allows the distributed system to carry out valuable tasks such as quantum key distribution (QKD), which guarantees secure communication through the laws of physics~\cite{bedington2017progress,bennett1992communication}.  Significant progress is currently being made towards the creation of a global quantum network~\cite{simon2017towards}, and it is becoming an increasing priority to find further applications that can be built on such a platform.

A blockchain is a type of classical database that contains records about the past, such as a history of financial (or other) transactions.  Its unique design~\cite{nakamoto2008bitcoin} makes it very difficult to tamper with, and it also does not require a centralised institution to maintain its ongoing accuracy. 
A notable result~\cite{aggarwal2017quantum, kiktenko2017quantum} is~that scalable quantum computers could successfully break the cryptographic protocols that are used to secure (classical) blockchains, as well as the digital security of the modern world.  With the advent of a quantum computing race~\cite{castelvecchi2017quantum,castelvecchi2018quantum}, there have been various proposals for modified classical blockchains to protect against such an attack~\cite{king2012ppcoin,witte2016blockchain,buterin2016chain}.  However, their reliability can be questioned, given the large research effort to find new quantum algorithms~\cite{mcmahon2007quantum,montanaro2016quantum, zeng2017first} which could potentially undermine such work.
In addition to this, classical blockchains with added quantum features have also been put forward~\cite{kalinin2018blockchain,jogenfors2016quantum,sapaev2018quantum, behera2018quantum,tessler2017bitcoin, ikeda2017qbitcoin}. One in particular~\cite{kiktenko2017quantum} adds a QKD network layer (which protects the relevant sub-algorithm against a quantum computing attack) to a classical blockchain.
A more desirable solution would be an intrinsically quantum blockchain, which is constructed out of quantum information, and~whose design is fully integrated into a quantum network.  This would provide the benefit of a QKD layer as well other potential quantum advantages over a classical blockchain.  

In this article, we propose a conceptual design for a quantum blockchain using entanglement in time~\cite{Lev,brukner2004quantum,ringbauer2018multi}.  Nonclassical correlations between temporally separated quantum systems have manifested themselves through various physical settings; the particular case used in this work involves entanglement in time between photons that do not simultaneously coexist~\cite{megidish2013entanglement}.  

Our novel methodology encodes a blockchain into these temporally entangled states, which can then be integrated into a quantum network for further useful operations.  We also show that entanglement in time, as opposed to entanglement in space, plays the pivotal role for the quantum benefit over a classical blockchain.  
As discussed below, all the subsystems of this design have already been shown to be experimentally realisable.  Furthermore, if such a quantum blockchain were to be constructed, we would show that it could be viewed as a quantum networked time machine.

\section{Classical Blockchain} 

The classical blockchain is composed of a blockchain data structure and a network consensus protocol; the former is the database, while the latter provides the decentralisation feature.  The aim of a blockchain is to have a single database of records about the past that every node in the network can agree on.  Furthermore, it should not require a centralised management node.  
It will be helpful to construct a physical model to describe this classical information system, i.e., its kinematic and dynamic~properties.  

We start with the kinematic features.  Records about the past, which occurred at around the same time, are received and collected into a data block.  These blocks are time-stamped to ensure that the data existed at the specified time.  Furthermore, the blocks are linked in chronological order through cryptographic hash functions~\cite{katz2014introduction}.

If an attacker tries to tamper with a particular block, these cryptographic hash functions can be used to ensure, with a high degree of confidence, the invalidation of all future blocks following the tampered block; this makes the classical blockchain extremely fragile, and thus sensitive to tampering.  Hence, the older the time stamp on the block, the more secure it is in the blockchain.  The key benefit to achieve {through this sensitivity} is that it should be very difficult to (successfully) tamper with a block without invalidating {the blockchain}.

Another way to achieve this benefit in the kinematic case is to have a large distributed network with each node hosting a {local} individual copy of the blockchain.  If a dishonest node tampers with its local copy, it does not affect the other copies in the other nodes of the network.

In the dynamic case, we want to examine how a blockchain lengthens over time.  The objective is to add valid blocks without a centralised institution; there are a number of ways this could be accomplished.
One current classical design does this by invoking a node on the network to confirm the validity of records in a new block, and then broadcasting that block to other nodes.  The~different nodes accept the block if they can successfully link it to their own copy of the blockchain through the cryptographic hash functions.  For this procedure to maintain ongoing accuracy, the~validating node gets chosen at random for each block; this prevents preplanned node-specific attacks.  Furthermore, the validating node  is also incentivised through the network for carrying out these tasks.  Despite some dishonest nodes, this is all achieved through consensus algorithms like proof-of-work or proof-of-stake~\cite{king2012ppcoin}. 


From our analysis, we see that the relevant performance benefits are nontampering, and also maintaining ongoing accuracy in a decentralised manner. We aim to show that a fully quantum blockchain, at an abstract level, could provide an advantage over classical blockchain on these performance metrics.

\begin{widetext}
\section{Quantum Blockchain}  

{In this section, our aim is to replace the data structure component of the classical blockchain with a quantum system.}  In quantum information theory, quantum systems are described as information carriers, with an encoding and decoding process. 
For the case of a blockchain, we capture the notion of the chain through the nonseparability (entanglement) of quantum systems (e.g., photons). For a bipartite system $\ket{\psi}_{AB}$, this means that:

\begin{equation}
\ket{\psi}_{AB} \ne \ket{a}_{A}\ket{b}_{B}, 
\end{equation}
for all single qubit states $\ket{a}$ and $\ket{b}$; the subscripts refer to the respective Hilbert spaces.
In particular, multipartite GHZ (Greenberger--Horne--Zeilinger) states~\cite{ghz,  carvacho2017experimental} are ones in which all subsystems contribute to the shared entangled property.  This enables us to create the concept of a chain.  

To create the appropriate code to utilise this chain, it is helpful to use a concept from superdense coding~\cite{bennett1992communication}.  In this protocol, a code converts classical information into spatially entangled Bell states; two classical bits, $xy$, where $xy = 00, 01, 10$ or $11$, are encoded to the state:

\begin{equation}
\ket{\beta_{xy}} = \frac{1}{\sqrt{2}}(\ket{0}\ket{y} + (-1)^{x}\ket{1}\ket{\bar{y}}), 
\end{equation}
where $\bar{y}$ is the negation of $y$.  Given that Bell states are orthonormal, they can be distinguished by quantum measurements.  This decoding process allows one to extract the classical bit string, $xy$, from~$\ket{\beta_{xy}}$.  

For our conceptual design, we temporarily simplified the data characterising the records in the classical block to a string of two bits.  Our encoding procedure converted each block with its classical record, say $r_{1}r_{2}$, into a temporal Bell state~\cite{megidish2013entanglement}, generated at a particular time, say $t=0$:

\begin{equation}
\ket{\beta_{r_{1}r_{2}}}^{0, \tau} = \frac{1}{\sqrt{2}}(\ket{0^{0}}\ket{r_{2}^{\tau}} + (-1)^{r_{1}}\ket{1^{0}}\ket{\bar{r_{2}}^{\tau}}).
\end{equation}

The superscripts in the kets signify the time at which the photon is absorbed; notice that the first photon of a block is absorbed immediately. For our purposes, this provides a way to do time stamps for each block.

{Physically, such temporal Bell states were experimentally generated in the work by the authors of \cite{megidish2013entanglement}.  In their procedure, spatially entangled qubits were represented through polarized photons:

\begin{equation}
\ket{\phi \pm} = \frac{1}{\sqrt{2}}(\ket{h_{a}h_{b}} \pm {\ket{v_{a}v_{b}}}), \quad \ket{\psi \pm} = \frac{1}{\sqrt{2}}(\ket{h_{a}v_{b}} \pm {\ket{v_{a}h_{b}}}),
\end{equation}       
where $h_{a}$ ($v_{a}$) represent the horizontal (vertical) polarization in spatial mode $a$ ($b$).  To create the temporally entangled states, consecutive pairs of spatially entangled pairs were generated at well-defined times separated by time interval $\tau$:

\begin{equation}
\ket{\psi -}_{a,b}^{0,0} \otimes \ket{\psi -}_{a,b}^{\tau,\tau} = \frac{1}{{2}}(\ket{h_{a}^{0} v_{b}^{0}} - {\ket{v_{a}^{0}h_{b}^{0}}}) \otimes (\ket{h_{a}^{\tau} v_{b}^{\tau}} - {\ket{v_{a}^{\tau}h_{b}^{\tau}}}),
\end{equation}
where the added superscripts provide the time labels for the photons.  In the experiment, a delay line of time $\tau$ is introduced to one of the photons of each entangled pair.  This resulting state equated to:

\begin{equation}
\ket{\psi -}_{a,b}^{0,\tau} \ket{\psi -}_{a,b}^{\tau,2\tau} = \frac{1}{2}(\ket{\psi +}_{a,b}^{0,2\tau} \ket{\psi +}_{a,b}^{\tau,\tau} - \ket{\psi -}_{a,b}^{0,2\tau} \ket{\psi -}_{a,b}^{\tau,\tau} - \ket{\phi +}_{a,b}^{0,2\tau} \ket{\phi +}_{a,b}^{\tau,\tau} + \ket{\phi -}_{a,b}^{0,2\tau} \ket{\phi -}_{a,b}^{\tau,\tau}).
\end{equation} 

When Bell projection was carried out on two photons at time $t= \tau$, entanglement was created between the photon absorbed at $t=0$ and the photon absorbed at $t=2\tau$; this is despite the fact that the latter two photons have never coexisted.}

{Going back to our design, as} records are generated, the system encodes them as blocks into temporal Bell states; these photons are then created and absorbed at their respective times.  A specific example of such blocks would be:
 
\begin{equation}
\ket{\beta_{00}}^{0, \tau}, \qquad \ket{\beta_{10}}^{\tau, 2\tau}, \qquad \ket{\beta_{11}}^{2\tau,3\tau},
\end{equation}
and so forth.  
To create the desired quantum design, the system should chain the bit strings of the Bell states together in chronological order, through an entanglement in time~\cite{Lev,brukner2004quantum,ringbauer2018multi}.  

Such a task can be accomplished using a fusion process~\cite{megidish2012resource} in which temporal Bell states are recursively projected into a growing temporal GHZ state.  {Physically, the fusion process is carried out through the entangled photon-pair source, a delay line, and a polarising beam splitter (PBS).  As an example, two Bell states can be fused into the following four-photon GHZ state:
\begin{eqnarray}
\ket{\psi +}_{a,b}^{0,0} \otimes \ket{\psi +}_{a,b}^{\tau,\tau} &\xrightarrow{delay} \ket{\psi +}_{a,b}^{0,\tau} \otimes \ket{\psi +}_{a,b}^{\tau,2\tau}  
= \frac{1}{{2}}(\ket{h_{a}^{0} v_{b}^{\tau}} + {\ket{v_{a}^{0}h_{b}^{\tau}}}) \otimes (\ket{h_{a}^{\tau} v_{b}^{2\tau}} + {\ket{v_{a}^{\tau}h_{b}^{2\tau}}}) \\ &\xrightarrow{PBS} \frac{1}{2}(\ket{h_{a}^{0} v_{b}^{\tau}v_{a}^{\tau}h_{b}^{2\tau}} + \ket{v_{a}^{0} h_{b}^{\tau}h_{a}^{\tau}v_{b}^{2\tau}}) = \ket{GHZ}^{0,\tau, \tau, 2\tau}.
\nonumber
\end{eqnarray}

In this GHZ state, entanglement exists between the four photons that propagate in different spatial modes and exist at different times.}

Implementing  this {procedure in our design}, the state of the quantum blockchain, at $t=n\tau$ (from~$t=0$) is given by:

\begin{eqnarray}
\label{QBLOCK}
&&
\ket{GHZ_{r_{1}r_{2}\ldots r_{2n}}}^{0,\tau, \tau, 2\tau, 2\tau \ldots , (n-1)\tau, (n-1)\tau, n\tau}  
= \frac{1}{\sqrt{2}}(\ket{0^{0}r_{2}^{\tau}r_{3}^{\tau}\ldots r_{2n}^{n\tau}} + (-1)^{r_{1}}\ket{1^{0}\bar{r}_{2}^{\tau}\bar{r}_{3}^{\tau}\ldots \bar{r}_{2n}^{n\tau}}). 
\end{eqnarray}

The subscripts on the left hand side  
of Equation (\ref{QBLOCK}) denote  the concatenated string of all the blocks, while superscripts refer to the time stamps.  The time stamps allow each blocks' bit string to be differentiated from the binary representation of the temporal GHZ basis state.  Note that at $t=n\tau$, there is only one photon~remaining.

The dynamics of this procedure can be illustrated with our example above.  Out of the first two blocks, $\ket{\beta_{00}}^{0, \tau}$ and $\ket{\beta_{10}}^{\tau, 2\tau}$, the system creates the (small) blockchain $\ket{GHZ_{0010}}^{0, \tau, \tau, 2\tau}$.  Concatenating the third block then produces  $\ket{GHZ_{001011}}^{0, \tau, \tau, 2\tau, 2\tau, 3\tau}$ .
The decoding process extracts the classical information, $r_{1}r_{2}\ldots r_{2n}$, from the state (Equation (\ref{QBLOCK})).  In recent work, it was shown how to characterise any such temporally generated GHZ state efficiently compared to standard tomography techniques.  This can be accomplished without measuring the full photon statistics, or even detecting them~\cite{megidish2017quantum}.
Each of the operations above have been explicitly shown to be experimentally realisable, at least in simple cases~\cite{megidish2012resource, megidish2013entanglement, megidish2017quantum}.

{The scalability of these temporally entangled systems was considered in Reference}~\cite{megidish2013entanglement}.  The key implication of their work is ,and we quote, ``any number of photons are generated with the same setup, solving the scalability problem caused by the previous need for extra resources. Consequently, entangled photon states of larger numbers than before are practically realizable.''  Our proposal certainly does not yet meet immediate engineering considerations for a quantum blockchain, but it shows a visible path forward {towards scalability}.

It is important to note that at this stage of development, we are advocating a conceptual mathematical design for a new quantum {information} technology.  It should be viewed as analogous to early quantum algorithms (Deutsch's algorithm~\cite{deutsch1985quantum}, Deutsch-Jozsa algorithms~\cite{deutsch1992rapid}{)}.  In the 1980s, the~engineering considerations for quantum computers were not taken into account.  In the 1990s, when~Shor's algorithm and Grover's algorithm were developed, the experimental realisation of quantum computers was almost seen as an impossible project.  Nonetheless, their work was certainly of interest to the wider community~\cite{Grover}.

\end{widetext}

\null
\vspace{2cm}
\clearpage

\section{Quantum Network}  

Recall that a classical blockchain system has a number of different components:  A blockchain {data structure}, a copy of {this} blockchain {data structure} at each node of {a classical} network, and a {consensus network} algorithm to verify the correctness of new blocks (before adding that new block to a blockchain).  

{In our design, we replaced the classical network with a quantum network}. {In addition}, digital signatures would be covered by a QKD protocol as stated in Reference~\cite{aggarwal2017quantum}.  In fact, others have used this way of reasoning when introducing new quantum protocols.  For example, in the paper that introduced the $\theta$-protocol~\cite{mccutcheon2016experimental}, the authors also simply assumed a QKD layer before moving onto their original work.  We quote from their paper, ``it is assumed that the verifier and each of the parties share a secure private channel for the communication. This can be achieved by using either a one-time pad or quantum key~distribution.''

{Furthermore}, each node on the quantum network would host a copy of the quantum blockchain~ (Equation (\ref{QBLOCK})); hence, if a node tampers with its own local copy, it does not affect the copies at the other nodes {analogous to the classical case}.   New blocks (that come from a sender) need to be verified for their correctness, before being copied and added to each node's blockchain.  Since correct blocks are GHZ entangled states, one needs a verification test to do it.
	
At this stage of the design, we assumed that newly generated blocks are spatial GHZ states (converting this to the related temporal case is at this stage of the design process unnecessary and is left for future work).  As in the classical case, the objective is to add valid blocks in a decentralised manner. The~challenge is that the network can consist of dishonest nodes, and the generated blocks can come from a dishonest source.  
To solve this problem, the quantum network uses the $\theta$-protocol~\cite{mccutcheon2016experimental}, which is a consensus algorithm where a random node in the {quantum} network can verify that the untrusted source created a valid block (spatial GHZ state).  More crucially, this is accomplished in a decentralised way using other network nodes, which may also be dishonest (Byzantine nodes). 

To start off {this verification protocol}, we needed to pick a randomly chosen verifier node (analogous to proof-of-stake or proof-of-work); this can be accomplished through a low level sub-algorithm involving a quantum random number generator.
The untrusted source shares a possible valid block, an $n$-qubit state. Since it knows the state, it can share as many copies of the block as is needed without running afoul of the no-cloning theorem.  For verification, it distributes each of the qubits to each node, $j$. 
The verifying node generates random angles $\theta_{j} \in \left[0, \pi \right) $ such that $\sum_{j}\theta_{j}$ is a multiple of $\pi$.  The~(classical) angles are distributed to each node, including the verifier. 

They respectively measure their qubit using the basis: 
%

\enlargethispage{30pt}
\begin{eqnarray}
\ket{+_{\theta_{j}}} = \frac{1}{\sqrt{2}}\left(\ket{0}+e^{i\theta_{j}}\ket{1}\right), 
\\
\ket{-_{\theta_{j}}} = \frac{1}{\sqrt{2}}\left(\ket{0}-e^{i\theta_{j}}\ket{1}\right).
\end{eqnarray}

The results, $Y_{j}=\{0,1\}$, are sent to the verifier. 
If the $n$-qubit state was a valid block, i.e., 
 a spatial GHZ state, the necessary condition:

\begin{equation}
\oplus_{j}Y_{j} = \frac{1}{\pi} \sum_{j}\theta_{j} \quad   (\text{mod 2}), 
\end{equation}
is satisfied with probability $1$. 

The protocol links the verification test to the state that is used; Reference~\cite{mccutcheon2016experimental} explicitly mentions this, and we quote, ``It is important to remark that our verification protocols go beyond merely detecting entanglement; they also link the outcome of the verification tests to the state that is actually used by the honest parties of the network with respect to their ideal target state. This is non trivial and of great importance in a realistic setting where such resources are subsequently used by the parties in distributed computation and communication applications executed over the network.''
Hence, the block can be copied and distributed to each node on the network to be added onto their blockchain.
This $\theta$-protocol has also been experimentally realised in simple cases, with quantitative analysis provided on cases involving dishonest nodes~\cite{mccutcheon2016experimental}.

\section{{Discussion}}  

For an analysis of the quantum benefit, we looked primarily at the blockchain's ability to be rendered tamper-proof. The~classical blockchain data structure operates through time-stamped blocks with cryptographic hash functions linking them in a chronological order. If an attacker tampers with a particular block, all future blocks following the tampered block are invalidated.  {This sensitivity to such disturbances is the mechanism which makes it difficult to tamper with a blockchain without invalidating it.}  

For the quantum blockchain, we replaced the important functionality of time-stamped blocks and hash functions linking them with a temporal GHZ state with an entanglement in time.  The quantum advantage is that {the sensitivity towards tampering is significantly amplified}, meaning that the full local copy of the blockchain is destroyed if one tampers with a single block (due to entanglement); on~a classical blockchain, only the blocks after the tampered block are destroyed (due to cryptographic hash functions), which leaves it open to vulnerabilities.  For the classical case, it is often stated that the farther back the block was time-stamped in, the more ``secure" it is; this is precisely because of the above invalidation.

More specifically, with just a spatial GHZ state, the measurement correlations of these states are stronger than what a classical system could ever produce.  In this spatial entanglement case, if an attacker tries to tamper with any photon, the full local copy of the blockchain would be invalidated immediately; this already provides a benefit over the classical case where only the future blocks of the tampered block are invalidated.

The temporal GHZ blockchain (Equation (\ref{QBLOCK})) adds a far greater benefit in that the attacker cannot even attempt to access the previous photons, since they no longer exist.  They can at best try to tamper with the last remaining photon, which would invalidate the full state.  Hence, in this application of quantum information, we see that the entanglement in time provides a far greater security benefit than an entanglement in space. 

There still needs to be a careful case-by-case analysis of potential tampering with the ultimately classical measurement results, but that would entail full security proofs, which is left for future work.  It~is important to distinguish between the classical case and quantum case like it was done for quantum algorithms with bounded probability for errors vs. classical deterministic algorithms. Showing such detailed security proofs would be a research paper in itself.

In terms of its network component, the classical blockchain's most important feature is time-stamped blocks with cryptographic hash functions linking them.  Without this, the concept of a blockchain simply does not exist.  However, with this concept, there are variations on various types of blockchain that can be considered.  Some are public and some are ``permissioned'' with the extreme version being centralised.  They have different consensus algorithms like proof of work or proof of stake.   Some have Byzantine fault tolerance and others do not.  
	
To be more specific, there are many well-known blockchains that do not provide Byzantine fault tolerance, such as earlier versions of hyperledger fabric~\cite{BFTLit}.  Hence, this sets the precedent that one does not necessarily need Byzantine fault tolerance to present a novel blockchain design.
Furthermore, it has been shown that Nakamoto's proof-of-work does not fulfil Byzantine fault tolerance.  That is,  no  formal proof  to this effect exists~\cite{PoW}; there is merely a probabilistic solution, and some argue that its probabilistic guarantees are not easy to relate to the consensus definitions given in traditional distributed computing literature~\cite{Dist1, Dist2}.  Given this unclear situation even in the classical case, it seems unreasonable to demand full security proofs for Byzantine fault tolerance in our quantum design.
Lastly, the $\theta$-protocol which we harness in our design has been analysed quantitatively on uses in cases that involve dishonest (Byzantine) nodes~\cite{mccutcheon2016experimental}.  
It is indeed correct to say that our consensus algorithm needs further low-level detail, but that would entail providing a low-level design; at this stage, it would not add much to the main work of introducing a quantum blockchain data structure.  Future work on quantum network consensus algorithms will build on this data structure.
	
The temporal GHZ state {that we used in our design} involves an entanglement between photons that do not share simultaneous coexistence, yet they share nonclassical measurement correlations.  
This~temporal nonlocality between two entangled photons that existed at different times was interpreted in Reference~\cite{megidish2013entanglement} as follows: ``\textit{...measuring the last photon affects the physical description of the first photon in the past, before it has even been measured. Thus, the ``spooky action" is steering the system's past}''.  
Stated more boldly, in our quantum blockchain, we can interpret our encoding procedure as linking 
the current records in a block, not to a record \textit{of} the past, but linking it to the actual record \textit{in} the past, a record which does not exist anymore. 

This research can be taken into various different directions.  To enhance the realistic possibility of this design being implemented, one should note that quantum networks are currently being realised on space-based satellite links~\cite{bedington2017progress}; therefore, spacetime effects need to be taken into account~\cite{bruschi2014spacetime, Bruschi:2013yeu}. 
This~would entail extending this work into the regime of relativistic quantum information.  We~speculate that a blockchain can then be encoded into a different temporally entangled system, namely, the entanglement between the future and past in the quantum vacuum~\cite{olson2011entanglement, olson2012extraction}.  A realistic experimental proposal~\cite{sabin2012extracting} suggests that it is possible to transfer this future past quantum correlation into qubits that do not simultaneously coexist, which is the resource needed for our current design.

For another direction, it is interesting to note that designs for classical blockchains have also advanced in various ways, some with nontrivial casual data~\cite{buterin2016chain}.  One can look at combining such systems with other temporal quantum concepts, such as the work in quantum casual structures~\cite{oreshkov2012quantum, brukner2014quantum, chaves2015information} {or alternative temporal-based GHZ states~\cite{nowakowski2017quantum,nowakowski2018entangled}}.  This may provide variety of qualitatively different designs for a quantum blockchain. 

An audacious direction of research stems from the view that at each node, our encoding procedure can be interpreted as influencing the past. With all such nodes connected through quantum channels, the blockchain can be viewed as a quantum networked time machine.  On~the theoretical front, the system design may be harnessed to invent other useful applications where the full network collectively influences the past in nonclassical ways; this may also lead a type of information--theoretic investigation into the nature of time~\cite{Lloyd:2010nt, Lloyd:2011zz}. Furthermore, unlike~general relativistic time \mbox{machines~\cite{visser1995lorentzian, Hawking:1991nk, Visser:2002ua, Visser:1993}, all the subcomponents} of this system have shown to be realisable~\cite{megidish2012resource, megidish2013entanglement, megidish2017quantum, mccutcheon2016experimental}; this~suggests the possibility to experimentally probe time travel paradoxes through quantum information.
{At the very least, such a system could lead to experimental probes of quantum causality~\cite{brukner2014quantum, ringbauer2016experimental}} .

\section{Conclusions} 

We have outlined a conceptual design for a quantum blockchain using entanglement in time.  Our~primary innovation was in encoding the blockchain into a temporal GHZ state.  Furthermore, it was shown that entanglement in time, as opposed to entanglement in space, provides the crucial quantum benefit. 

This paper is meant to serve as a conceptual design for a new quantum {information} technology.  As such, a lot of detailed low level design gaps do exist, but the intention was to open up a novel area where at least the core functionalities are covered.  In the most cited quantum computing textbook \cite{nielsen2010quantum}, it is mentioned that the primitive ``Deutsch-Jozsa algorithm contains the seeds for more impressive quantum algorithms.''  Similarly, we believe that our conceptual quantum blockchain would be the seed for many detailed quantum blockchains that will build on our work.  Given the rise of classical blockchains and the realistic development of a global quantum network, this work can potentially open the door to a new research frontier in quantum information science.

\medskip
\noindent{\bf Author contributions:} Conceptualisation, D.R. {and M.V.}; Formal analysis, D.R. and M.V.; Funding acquisition, M.V.; Investigation, DR.; Methodology, M.V.; Project administration, M.V.; Supervision, M.V.

\medskip
\noindent{\bf Funding} D.R. was indirectly supported by the Marsden fund, administered by the Royal Society of New Zealand. 
M.V. was directly supported by the Marsden fund, administered by the Royal Society of New Zealand.

\medskip
\noindent{\bf Conflicts of interest:} The authors declare no conflict of interest.


\end{document}